\documentclass[a4paper,twocolumn,11pt,accepted=2017-05-09]{quantumarticle}
\pdfoutput=1
\usepackage[utf8]{inputenc}
\usepackage[english]{babel}
\usepackage[T1]{fontenc}
\usepackage{amsmath}
\usepackage{hyperref}
\usepackage{tabularray}
\usepackage{booktabs}
\usepackage{float}
\usepackage{multirow}

\usepackage{tikz}
\usepackage{lipsum}

\begin{document}

\title{Study of the Influence of Implant Material on Magnetocardiography Measurements Using SQUID Sensors}

\author{Ho-Seong Lee}
\affiliation{Department of Biomedical Engineering, Korea University, 46 Gaeunsa 2-gil, Seongbuk-gu, Seoul, 02842, Korea}
\affiliation{AMCG, 14F, 331, Gangnam-daero, Seocho-gu, Seoul, 06627, Korea}
\email{hoseonglee14@gmail.com}
\orcid{0000-0002-2445-2701}
\author{Jae-Hyun Ahn}
\orcid{0009-0001-5154-824X}
\affiliation{AMCG, 14F, 331, Gangnam-daero, Seocho-gu, Seoul, 06627, Korea}
\author{Yong-Hwan Kim}
\affiliation{AMCG, 14F, 331, Gangnam-daero, Seocho-gu, Seoul, 06627, Korea}
\orcid{0000-0002-1087-5615}
\maketitle

\begin{abstract}
  Magnetocardiography (MCG) system is a medical device that diagnoses cardiac disease by measuring magnetic fields generated from electric currents flowing through the myocardium. However, the accuracy of measurement data can be degraded if strong magnetic materials are present or magnetic field changes occur near the MCG system. With the widespread use of implants, the number of patients with metallic implants is increasing, but there is a lack of in-depth research on the potential impact of implant materials on the results of the MCG examination. This study aims to analyze the effect of implant materials on MCG measurements and establish relevant criteria.\\
  In this study, a 96-channel MCG system employing Superconducting Quantum Interference Device (SQUID) sensors, specifically first-order gradiometers based on the Double Relaxation Oscillation SQUID (DROS) method, and a Magnetically Shielded Room (MSR) were utilized. Titanium-6Aluminum-4Vanadium ELI (Ti-6Al-4V ELI, ASTM F136) was selected as the representative implant material sample. Experiments were conducted under extreme conditions, where a sample significantly larger than an actual implant was placed as close as possible to the sensors.\\
  As a result, when the implant material was at the minimum distance to the sensor, the noise increase was approximately 0.7 fT/$\sqrt{\mathrm{Hz}}$, which satisfies the sensitivity criteria for MCG. Furthermore, since these results were obtained under severely adverse conditions designed to maximize the noise impact, it is anticipated that the effect would be even more negligible in actual clinical settings.\\
  In conclusion, it was confirmed that common implant materials, such as Ti-6Al-4V ELI, have little to no effect on MCG measurements. However, as the experiments were not conducted with the material inserted into the human body, unlike actual clinical environments, the generation of magnetic fields due to micromotion has not been verified, thus requiring further experimentation.
\end{abstract}

\begin{figure}[h]
    \centering
    \includegraphics[width=1\linewidth]{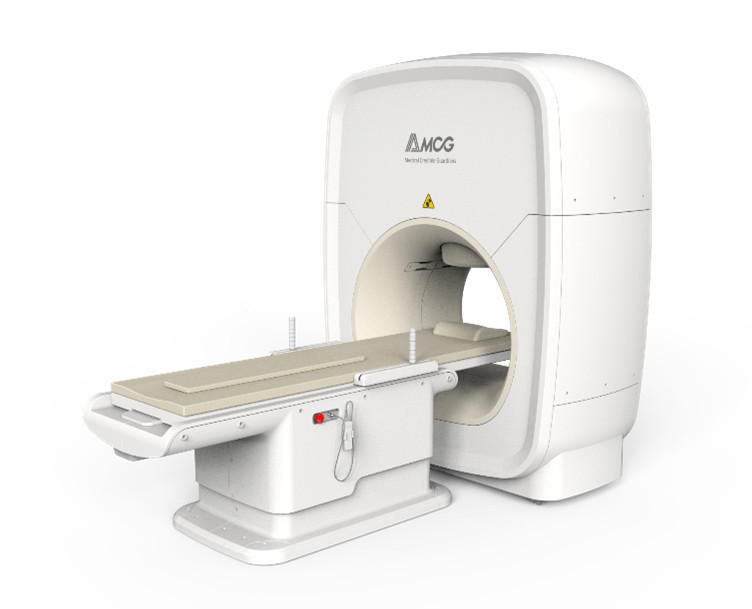}
    \caption{Magnetocardiography system}
    \label{fig:1}
\end{figure}

\section{Introduction}
Magnetocardiography (MCG) is a medical device that diagnoses heart diseases by measuring magnetic fields generated by electric currents flowing through the heart muscle. It is a very low-risk examination method for the human body as it involves no radiation exposure, no contrast agents, and allows for non-contact, non-invasive examinations.\cite{Magnetocardiographyforthedetectionofmyocardialischemia, Physicalprinciplesandinstrumentation, Clinicalapplications, Clinicalapplicationofmagnetocardiography} However, if a strong magnetic object is near the MCG system or unexpected magnetic field changes occur during the examination, the accuracy of the measured data can decrease. For this reason, MCG examinations are typically performed within a Magnetically Shielded Room (MSR) to minimize magnetic field interference.\cite{ReviewofmagnetocardiographytechnologybasedonSQUID}\\
The MCG system used in this study employed Superconducting Quantum Interference Device (SQUID) sensors. Most SQUID sensors used in MCG utilize low-temperature superconducting Nb Josephson junctions for DC-SQUID, which often suffer from insufficient Signal-to-Noise Ratio (SNR) and the disadvantage of requiring calibrations for optimal feedback resistance-coils after SQUID characteristic evaluation. To overcome these limitations, the SQUID sensors used in this study adopted the Double Relaxation Oscillation SQUID (DROS) method. DROS offers an advantage over the DC-SQUID method by achieving an approximate 10-fold increase in the transduction coefficient, which simplifies the front-end amplifier configuration, making it a more suitable method for multichannel MCG systems.\cite{ReviewofmagnetocardiographytechnologybasedonSQUID}\\
There is always a possibility of noise generation from magnetic materials inside or outside the MSR during MCG examinations. Observable noise sources in MCG measurements using SQUID sensors include vibration from nearby magnetic materials, DC/AC external magnetic fields, and hardware circuit noise. Among these, the noise most significantly affected by internal magnetic materials is from the vibration of the magnetic material itself.\cite{Superconductingquantuminterferencedeviceinstrumentsandapplications}\\
With advances in medical technology, the number of patients with implants, such as artificial joints and dental prostheses, which can be inserted into the human body, is increasing.\cite{A15yearretrospectiveanalysis, AnalysisofdifferencesaccordingtoDentalImplantMaterial} However, for patients with metallic implants, it is difficult to accurately know the degree of magnetism of the implant material. Furthermore, the removal of implants for MCG examinations is often impractical, making it impossible to rule out the possibility of their impact on examination results.\\
Previous research on diagnostic medical devices has consistently reported studies that measure physical properties of metallic materials inserted into the human body, such as permeability, magnetic susceptibility, and electrical conductivity, and evaluate their effects on image artifacts and noise.\cite{MagneticsusceptibilityandelectricalconductivityofmetallicdentalmaterialsandtheirimpactonMRimagingartifacts, Effectivemagneticsusceptibilityof3Dprintedporousmetalscaffolds} However, there has been no in-depth research analyzing the specific impact of implant materials on MCG measurements.\\
Therefore, this study aims to analyze the influence of commonly used implant materials on MCG measurements and to establish relevant criteria.

\section{Materials and Methods}
\begin{figure}[h]
    \centering
    \includegraphics[width=1\linewidth]{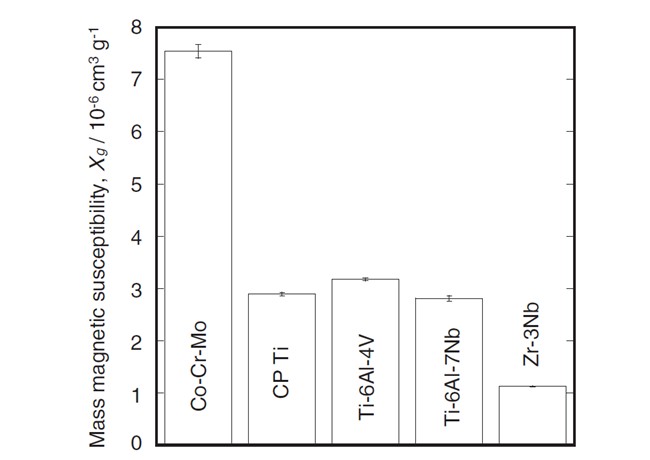}
    \caption{Magnetic susceptibilities of biomaterials used for medical devices.\cite{EffectsofphaseconstitutionofZrNballoysontheirmagneticsusceptibilities}}
    \label{fig:2}
\end{figure}

\paragraph{Materials}
The MCG system (MCG-S, AMCG Co., Ltd., Seoul, South Korea) used in the experiment employed 96-channel SQUID sensors, and the experiment was conducted inside a Magnetically Shielded Room (MSR). Fig.~\ref{fig:1} The SQUID sensors were first-order gradiometers of the Double Relaxation Oscillation SQUID (DROS) type.\\
Titanium and its alloys are the most commonly used metallic biomaterials for implants.\cite{Comparativestudyofbiocompatibletitaniumalloyscontainingnontoxicelementsfororthopedicimplants} Among these, the Ti-6Al-4V alloy exhibits the highest magnetic susceptibility.\cite{EffectsofphaseconstitutionofZrNballoysontheirmagneticsusceptibilities} Fig.~\ref{fig:2} If Ti-6Al-4V alloy is confirmed to have no significant noise impact during MCG examinations using SQUID sensors, then the noise impact of materials with lower magnetic susceptibility can be guaranteed. Therefore, Ti-6Al-4V ELI, as defined in ASTM F136\cite{ASTMF136}, was selected as the sample for this study.\\
To obtain the most uniform signal from the SQUID sensors of the MCG, a Ti alloy sample large enough to sufficiently cover the sensor array is required. The MCG's sensor array used in the experiment measures 250mm (W) X 250mm (L). A Ti-6Al-4V ELI sample (Supra Alloys, Inc., CA, USA) with dimensions of 300mm (W) X 400mm (L) X 7.93mm (H) was used, which is larger than the sensor array.\\
Considering that the femur, the longest human bone, averages 459mm in length, the Ti-6Al-4V ELI sample used in this experiment is larger than the size and volume of most implants. Therefore, it can be considered a representative sample size for evaluating implants.\cite{EstablishmentofFractureCriteriaonHumanFemur, Simulationbaseddesignoforthopedictraumaimplants}

\paragraph{Environmental Conditions}
The average shielding effectiveness of the MCG system used in the experiment, as presented by the manufacturer, is 36dB ±10

\begin{figure}[h]
    \centering
    \includegraphics[width=1\linewidth]{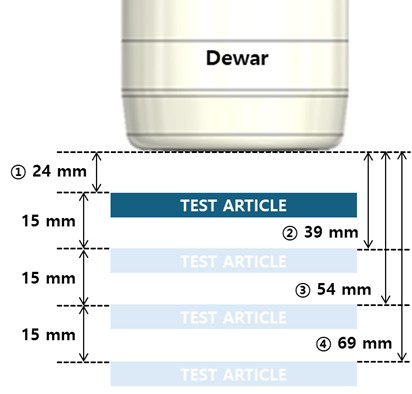}
    \caption{Positioning of test article}
    \label{fig:3}
\end{figure}

\paragraph{Test Procedure}
To investigate the magnetic influence the implant has depending on its distance to the MCG dewar, magnetic noise is measured first without a sample. Then, the gap between the bottom surface of the dewar and the top surface of the sample is adjusted to measure the magnetic noise. If the sample is placed directly on the bottom of the dewar (d = 0), vibrations caused by contact between the sample and the Outer Vessel can lead to difficulties in data interpretation. In order to evaluate the material's influence under identical conditions without other external factors, the initial position of the test sample is set 24 mm away from the bottom of the dewar. The test sample is positioned vertically using non-metallic jigs, each 15mm tall, as shown in Fig.~\ref{fig:3}. Before the experiment, the precise centering of the test sample within the dewar and its intended height are verified using a digital caliper.\\
When repeated 5 times for 1 minute at 5 Hz, a total of 1500 vibrations can be observed. This is sufficient to observe changes in signal magnitude, so each measurement is taken for 5 minutes.

\paragraph{Acceptance Criteria}
Magnetic fields generated by the heart typically range from 50 to 100 pT in magnitude. The MCG system used in the experiment is capable of measuring down to single digit fT, which is sufficient for measuring cardiac magnetic fields. The manufacturer's specified acceptance criteria for the noise level is 10 fT/$\sqrt{\mathrm{Hz}}$ @100 Hz, and this criterion is used to establish the acceptance criteria (AC) for this study.\cite{RecordingofCardiacExcitationUsingaNovelMagnetocardiographySystemwithMagnetoresistiveSensorsOutsideaMagneticShieldedRoom} However, since environmental noise can affect the evaluation of the material's intrinsic properties, if there is significant variability in measurements at 100 Hz due to external environmental factors, a frequency band as close as possible to 100 Hz is selected.\\
The results of each test must follow a normal distribution (P-value > 0.05). To confirm the reproducibility of the test, the average of the results from two sets of ten repetitions each should not be statistically significant (2-sample t-test, P-value > 0.05).

\section{Results and Discussion}
\begin{figure}[h]
    \centering
    \includegraphics[width=1\linewidth]{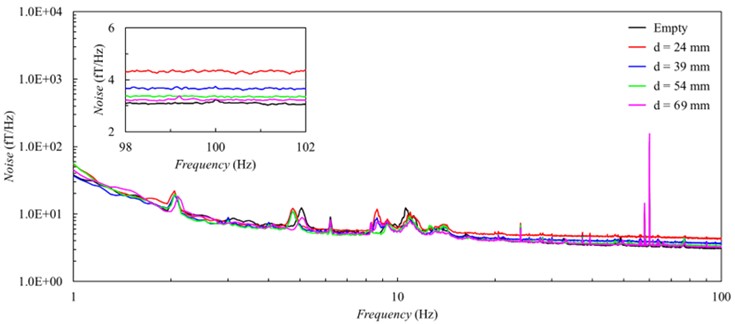}
    \caption{Comparison of the 10-time averaged noise in the frequency domain based on the distance between the Dewar and material, (Inset) variation in baseline height near 100 Hz for each distance.}
    \label{fig:4}
\end{figure}

\paragraph{Test Results}
Fig.~\ref{fig:4} illustrates the noise magnitude as a Power Spectral Density (PSD) graph according to the distance between the dewar and the material. Signals appearing as peaks at specific frequencies within the PSD represent environmental noise. The causes are diverse, including external equipment, electrical environments, and vibrations. Specifically, noise due to vibrations can be amplified by nearby magnetic materials, so the influence can be determined by observing changes in noise magnitude caused by magnetic materials.\\
However, in the noise measurement test using the Ti-6Al-4V ELI test material, such tendency was not observed. The environmental impact appears to be greater than the material's influence, which can be confirmed by comparing the empty state and d=69 mm in the 5 Hz band of Fig.~\ref{fig:4}.

\begin{figure}[h]
    \centering
    \includegraphics[width=1\linewidth]{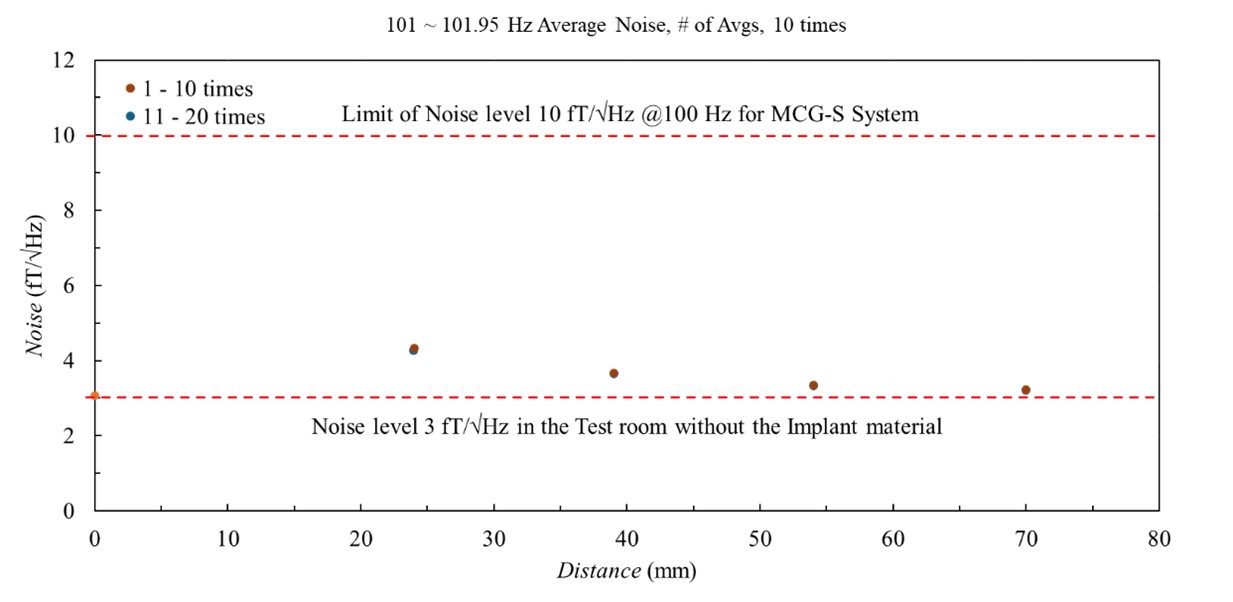}
    \caption{Comparison of the 10-time average noise at 100 Hz for each distance between the Dewar and material (brown) - First trial, (blue) Second trial}
    \label{fig:5}
\end{figure}

Nevertheless, a tendency for the baseline height of the noise spectrum to increase with the distance between the dewar and the material is observed. This is shown in Fig.~\ref{fig:4} (Inset). Since it is necessary to specify the height of the baseline in each measurement, the height at the 100 Hz band, which is the sensitivity standard presented by the MCG manufacturer, was calculated. As environmental noise occasionally occurs at 100 Hz, an average was taken on the noise values from 101 to 101.95 Hz was used for comparison to evaluate the intrinsic characteristics of the material. It was confirmed that the results across the entire frequency band near 100 Hz were lower than 10 fT/$\sqrt{\mathrm{Hz}}$, the sensitivity test criterion.\\
Fig.~\ref{fig:5} shows the average values of the noise baseline height as a function of the distance, and Table ~\ref{tab:table1} displays the average noise values from each trial. We can observe a tendency for the noise baseline to increase as the material approaches the dewar. In Fig.~\ref{fig:5}, the baseline height without the material is approximately 3 fT/$\sqrt{\mathrm{Hz}}$, while the MCG sensitivity criterion defined by the manufacturer is 10 fT/$\sqrt{\mathrm{Hz}}$. The baseline height at the closest proximity of 24 mm is approximately 4.3 fT/$\sqrt{\mathrm{Hz}}$, which satisfies the MCG-S sensitivity criterion.\\

\onecolumn

\begin{table*}[h]
    \centering

    \begin{tabular}{|c|c|c|c|c|c|}
        \hline
        Distance & Empty setup & 24 mm & 39 mm & 54 mm & 69 mm \\
        \hline
        T1   & 3.0791 & 4.3005 & 3.7099 & 3.3562 & 3.2291 \\\hline
        T2   & 3.0443 & 4.3427 & 3.6527 & 3.8490 & 3.2625 \\\hline
        T3   & 3.0473 & 4.3561 & 3.6097 & 3.3898 & 3.2183 \\\hline
        T4   & 3.1172 & 4.3297 & 3.7011 & 3.3820 & 3.2690 \\\hline
        T5   & 3.0319 & 4.2504 & 3.6398 & 3.3596 & 3.2622 \\\hline
        T6   & 3.0319 & 4.2809 & 3.6196 & 3.3540 & 3.2630 \\\hline
        T7   & 3.1262 & 4.3053 & 3.6453 & 3.3553 & 3.2504 \\\hline
        T8   & 3.0070 & 4.2947 & 3.6249 & 3.3632 & 3.2591 \\\hline
        T9   & 3.0492 & 4.3083 & 3.6394 & 3.3595 & 3.2498 \\\hline
        T10  & 3.0729 & 4.3148 & 3.7071 & 3.3857 & 3.2320 \\\hline
        T11  & 3.0478 & 4.3702 & 3.7016 & 3.3356 & 3.2420 \\\hline
        T12  & 3.0719 & 4.3575 & 3.7082 & 3.3787 & 3.2440 \\\hline
        T13  & 3.0570 & 4.3371 & 3.7061 & 3.3857 & 3.2515 \\\hline
        T14  & 2.9955 & 4.3750 & 3.7099 & 3.3893 & 3.2510 \\\hline
        T15  & 3.0460 & 4.3124 & 3.6508 & 3.3652 & 3.2397 \\\hline
        T16  & 3.0983 & 4.3869 & 3.6729 & 3.3927 & 3.2472 \\\hline
        T17  & 3.0919 & 4.3528 & 3.6674 & 3.3820 & 3.2498 \\\hline
        T18  & 3.0957 & 4.3664 & 3.6724 & 3.3862 & 3.2473 \\\hline
        T19  & 2.9079 & 4.3209 & 3.6595 & 3.3631 & 3.2434 \\\hline
        T20  & 3.1118 & 4.2807 & 3.6590 & 3.3684 & 3.2519 \\
        \hline
    \end{tabular}
    \caption{Noise Level@100 Hz (Unit:fT/$\sqrt{\mathrm{Hz}}$)}
    \label{tab:table1}
\end{table*}

\twocolumn

\begin{figure}[h]
    \centering
    \includegraphics[width=1\linewidth]{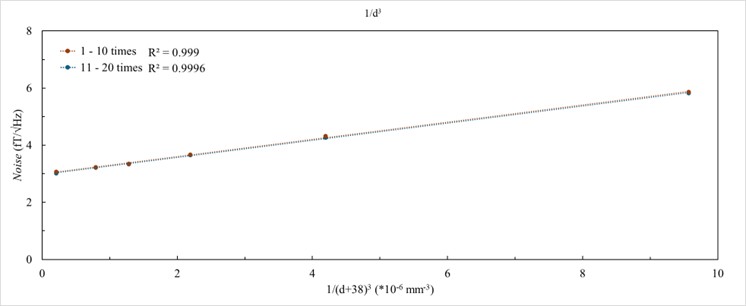}
    \caption{Relationship between the Dewar and the material distance (d) and noise baseline}
    \label{fig:6}
\end{figure}

Based on the results in Fig.~\ref{fig:5}, to predict the baseline height when the material is closest to the dewar, the trend between distance d and the baseline height is shown in Fig.~\ref{fig:6}.\\
Since the SQUID's pickup coil is located 38 mm from the bottom of the outer case, 38 mm was added to the measured distance d. The tendency for the baseline height to linearly proportion to $(d+38)^{-3}$ can be expressed as follows.

\begin{align}[h]
  \label{baseline}
  I = \alpha_i (d + 38)^{-3} + \beta_i \quad (i = 1, 2)   
\end{align}

\begin{figure}[h]
    \centering
    \includegraphics[width=1\linewidth]{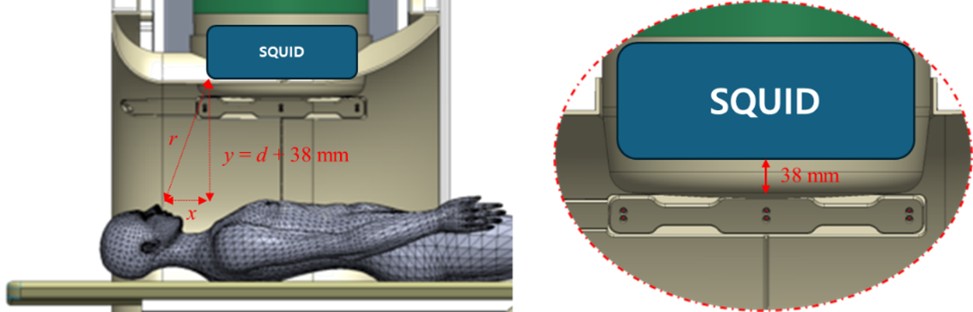}
    \caption{Diagram for calculating the distance r between the SQUIDs and the dental implant}
    \label{fig:8}
\end{figure}

\begin{figure}[h]
    \centering
    \includegraphics[width=1\linewidth]{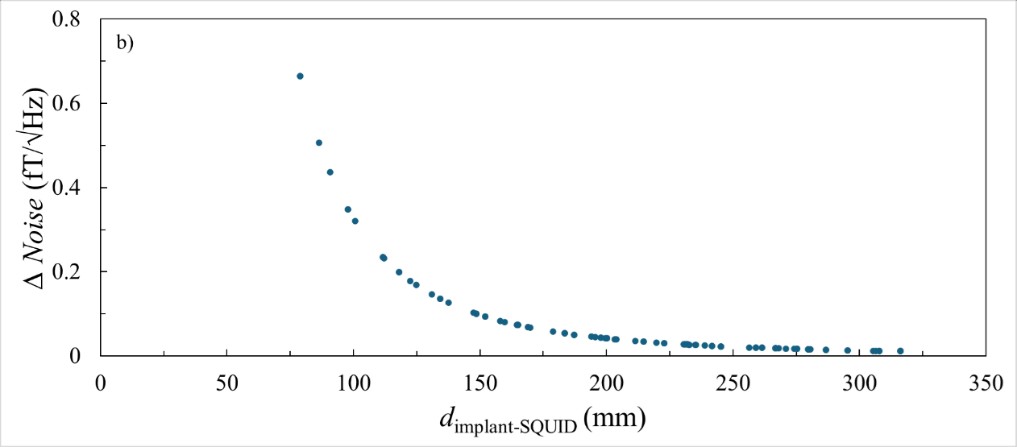}
    \caption{Changes in noise levels at 100 Hz as a function of the distance between each SQUID sensor and the material located at the (0, 190) coordinate.}
    \label{fig:9}
\end{figure}

$\alpha$ is the slope of the trendline, $\beta$ is the y-intercept of the trendline, and i represents the 1st and 2nd trials, respectively. For each trial, $\alpha$ values are (326339, 310817) and $\beta$ values are (2.951, 1.956). Through this, the baseline height at the minimum proximity distance (d=0) of the material can be extrapolated. If we assume the maximum proximity distance between the material and the dewar to be 0 mm, the maximum possible proximity distance between the material and the SQUID Sensor, based on the diagram, is 38 mm. The estimated baseline height at a distance of 38 mm can be calculated as (8.9, 8.6) fT/$\sqrt{\mathrm{Hz}}$ for each trial, respectively, and this also satisfies the MCG's sensitivity criteria.\\
\begin{figure}[h]
    \centering
    \includegraphics[width=1\linewidth]{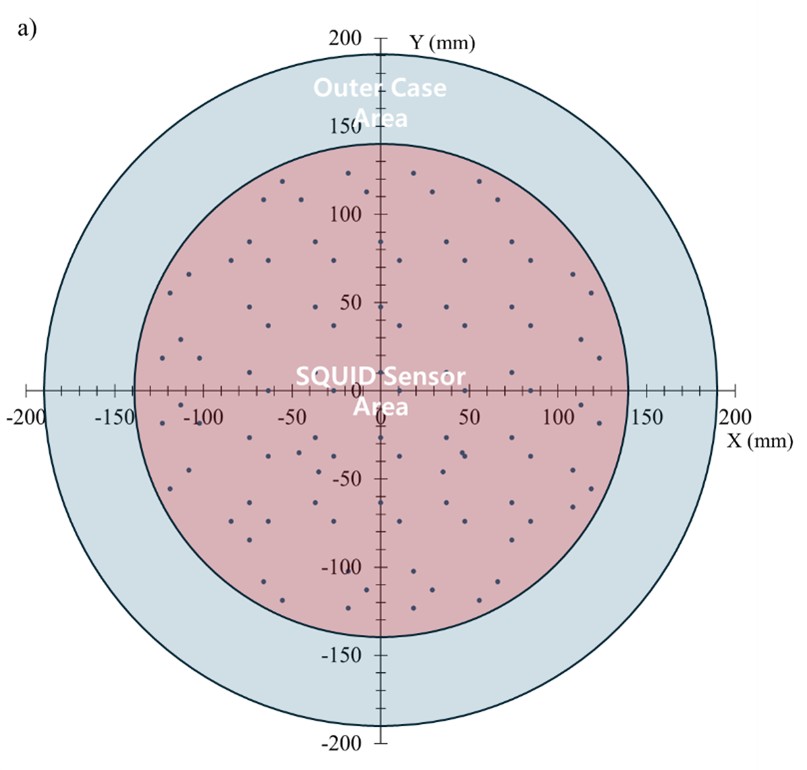}
    \caption{Planar distribution of SQUID sensors installed inside the Dewar (black dots) and their coverage area (red), along with the region physically inaccessible due to the outer case (blue)}
    \label{fig:7}
\end{figure}
In the case of a real patient, the closest and most commonly implanted device is likely to be a dental implant. The distance between a dental implant and the closest SQUID was estimated as follows: Fig.~\ref{fig:7} illustrates the area where SQUID sensors are distributed (red), the area physically inaccessible due to the outer case (blue), and the distribution of SQUID sensor locations (black dots). Assuming a person is positioned perpendicular to the Y-axis, the closest possible location for a dental implant is assumed to be near the outer boundary of the (blue) area. At this point, assuming the implant and SQUID sensor positions as shown in Fig.~\ref{fig:8}, with the chest against the dewar (d = 0) and the implant material at coordinates (0,190), the coordinates of the closest SQUID sensor are (18.5, 123.317). The distance calculated from this is approximately 78 mm. Since a distance of 78 mm already falls within the range of the performance test protocol, it is expected that implants made of Ti-4Al-6V will satisfy the MCG's sensitivity criteria, consistent with the conclusions above. Fig.~\ref{fig:9} shows the change in affected noise, calculated by determining the distances between the 96 MCG sensors and the material.\\

Disregarding jaw length, when considering the material's distance as the minimum possible distance to the Outer case region, the noise increase was approximately 0.7 fT/$\sqrt{\mathrm{Hz}}$. This suggests that it is not significantly affected even at the closest proximity.\\
This experiment assumed severe conditions designed to maximize the material's influence, using a sample larger in area and mass than typical implant specifications, and was conducted at a closer distance than the minimum possible proximity for actual implants. Consequently, the actual impact could be smaller than the results of this test.

\paragraph{Reproducibility and Distribution}
To minimize environmental variables affecting the evaluation of the sample's inherent properties, ten repeated measurements were performed. Data analysis was conducted using magnetic noise datasets in the frequency band near 100 Hz that passed both the 2-sample t-test and normality test. To confirm the reproducibility of this test, the 10 measurements were repeated again. As a result of conducting the 2-sample t-test and normality test, all test conditions showed a p-value greater than 0.05. This confirmed that each set of 10 measurements taken at different times followed a normal distribution and that the two tests performed at different times were identical. These results are recorded in Table ~\ref{tab:table2}.

\begin{table}[h]
\centering

\begin{tblr}{
  colspec = {|c|c|c|},
  hlines,
  vlines,
  row{1-2} = {valign=m, halign=c}, 
  row{3-Z} = {valign=m},          
  cell{1}{1} = {r=2}{halign=c,valign=m},
}
Data Set & Normality Test & {2 Sample T-Test} \\
& P value        & P value           \\
 
Empty -- T1 & 0.052 & \SetCell[r=2]{c,m} 0.238 \\
Empty -- T2 & 0.656 &                          \\
 
24 mm -- T1 & 0.149 & \SetCell[r=2]{c,m} 0.481 \\
24 mm -- T2 & 0.051 &                          \\
 
39 mm -- T1 & 0.419 & \SetCell[r=2]{c,m} 0.135 \\
39 mm -- T2 & 0.326 &                          \\
 
54 mm -- T1 & 0.552 & \SetCell[r=2]{c,m} 0.260 \\
54 mm -- T2 & 0.421 &                          \\
 
69 mm -- T1 & 0.053 & \SetCell[r=2]{c,m} 0.367 \\
69 mm -- T2 & 0.459 &                          \\
\end{tblr}
\caption{Normality Test and 2 Sample T-Test}
\label{tab:table2}
\end{table}

\section{Conclusion}
The Ti-6Al-4V ELI material used in the experiment did not affect the MCG measurement results. Therefore, it is can be concluded that implants made from materials with magnetic susceptibility equal to or lower than that of Ti-6Al-4V ELI are unlikely to influence MCG measurement outcomes.\\
However, since not all materials used in implants were tested, this conclusion could only be verified indirectly. Additionally, as the experiment was not conducted with the material implanted in the human body as it would be in a real clinical environment, the generation of magnetic fields due to micromotion has not been verified, thus necessitating further experimentation.

\bibliographystyle{quantum}

\end{document}